\documentclass[mathleft
]{an}
\usepackage{graphicx}
\usepackage{times}
\begin{document}

% The following seven commands are intended for editorial usage and should be ignored by
% the author(s).
\Pagespan{789}{}% Document's page range. 
% If second parameter is left empty, the last page is computed automatically.
\Yearpublication{2006}%
\Yearsubmission{2005}%
\Month{11}%   
\Volume{999}%  
\Issue{88}% 
% \DOI{This.is/not.aDOI}% 

\title{General Relativity effects and line emission}

\author{Giorgio Matt \thanks{
  \email{matt@fis.uniroma3.it}\newline}
%Example 
%for footnote, note the usage of the \texttt{fnmp}
%command as separator between institute number and footnote mark} 
}
\titlerunning{General Relativity effects and line emission}
\authorrunning{Giorgio Matt}
\institute{Dipartimento di Fisica, Universit\`a Roma Tre, Via della Vasca Navale 84,
I-00146 Roma, Italy }

\received{30 May 2005}
\accepted{11 Nov 2005}
\publonline{later}

\keywords{  accretion, accretion disks --  relativity -- line: profiles -- 
galaxies: active --  X-rays: binaries}

\abstract{General Relativity effects (gravitational redshift, light bending, ...) 
strongly modify the characteristics
of the  lines emitted close to the Black Hole in Active Galactic Nuclei
and Galactic Black Hole systems. These effects are reviewed and illustrated, 
with particular emphasis on line emission from the accretion disc. 
Methods, based on the iron line, to measure the two astrophysically relevant
 parameters of a Black Hole, the mass and spin, are briefly discussed.  
}

\maketitle

\section{Introduction}

With the advent of X-ray missions carrying on-board 
high sensitivity, moderate energy resolution instruments
(the first of which being ASCA, followed by $Beppo$SAX, $Chandra$, XMM--$Newton$ and
now $Suzaku$),
probing General Relativity (GR) effects on iron emission lines has become a reality, and it
is now an important part of the studies on Active Galactic Nuclei (AGN) and Galactic
Black Hole systems (GBH).\footnote{Relativistically distorted iron lines from accretion discs
around neutron stars have also possibly been observed, e.g. Di Salvo et al. (2005).}

In this paper I will review the main GR effects, 
and I will discuss how they can be used
to measure the two astrophysically relevant
 parameters of a Black Hole, the mass and the angular momentum (``spin''). The paper is
organized as follows: in Sec.~1 the basic concepts concerning Black Holes, of relevance
for understanding GR effects on line emission, are summarized. In Sec.~2  I will
discuss line emission from a relativistic accretion disc, while Sec.~3 is devoted to the
discussion of the strenghts and weaknesses of 
methods, based on iron emission lines, to measure the mass and spin of the
Black Hole. Conclusions are given in Sec.~4.

For a complete treatment of Black Holes  
the reader is deferred to standard textbooks like e.g. the ones
by Misner et al. (1973) and Chandrasekhar (1983), or to classical papers
like Bardeen et al. (1972). See also Fabian et al. (2000) and Reynolds \& Nowak (2003) for  
reviews on the iron line properties from relativistic accretion discs.

\section{Black Holes}

\begin{figure*}
\includegraphics[width=160mm,height=160mm]{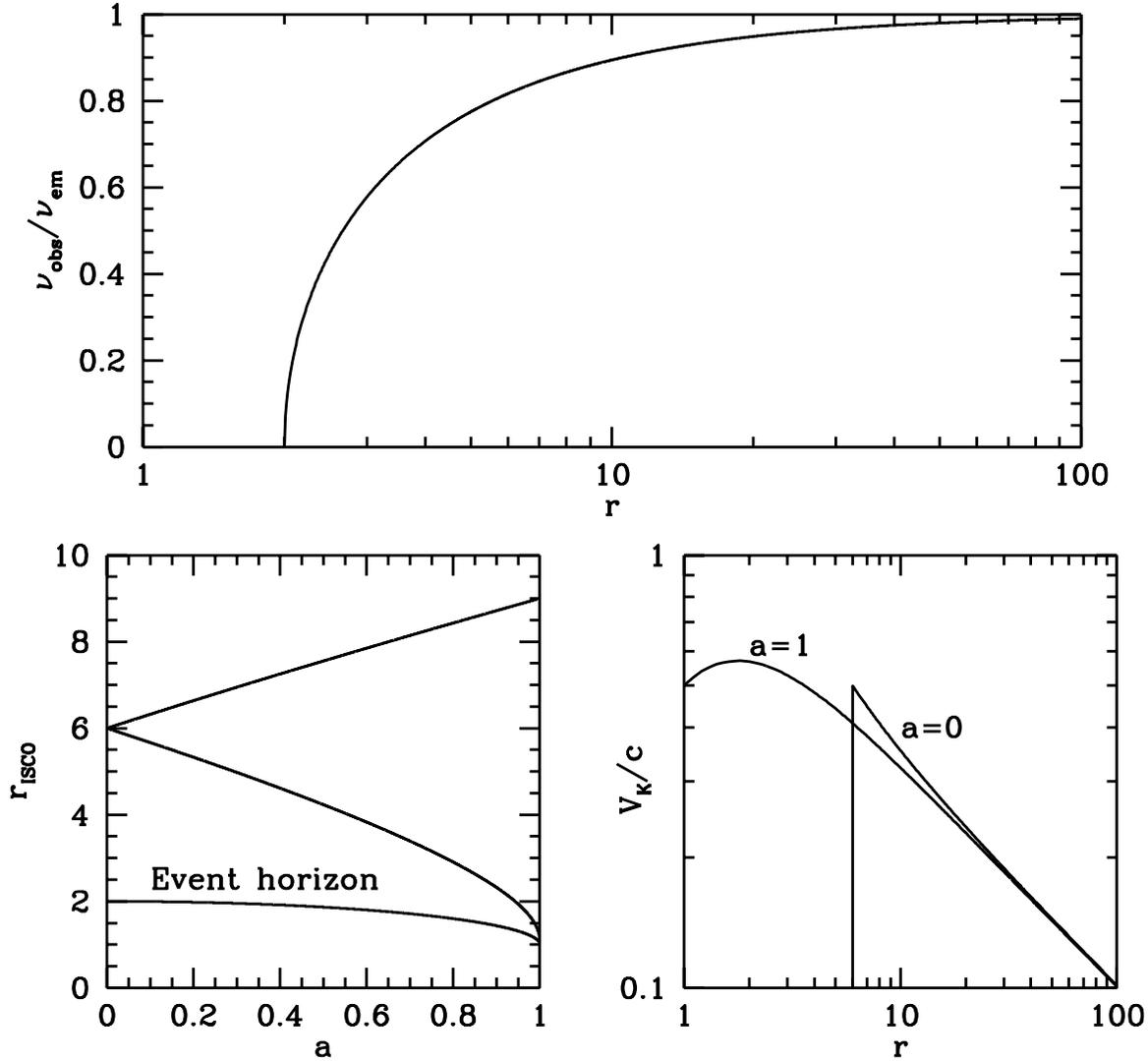}
\caption{Upper panel: the ratio between the observed and emitted
frequency of a photon (``gravitational redshift'') in Schwarzschild metric 
as a function of the radius at which the photon is emitted. Lower-left
panel: the radius of the Innermost Stable Circular Orbit as a function of
the BH spin. The lower (upper) curve refers to a co-- (counter--) rotating
disc. The radius of the Event Horizon is also shown for comparison. Lower-right
panel: the Keplerian velocity (in the LNRF) in the equatorial plane as a 
function of radius, for a static and a maximally rotating BH.}
\label{all}
\end{figure*}

A Black Hole (BH) or, better, the space-time around it, is fully characterized
by only three quantities: its mass $M$, angular momentum $J$ 
and electric charge $Q$.
The latter is usually assumed to be negligible for astrophysically relevant
Black Holes. The space time around a BH is described by the Kerr-Newman metric which,
when $Q$=0, reduces to the slightly simpler Kerr metric. If also $J$ 
is null, than the metric is the much simpler Schwarzschild one. It may be interesting
to remember that, while the static solution was found by Karl Schwarzschild
in 1916 (i.e., only one year after the publication by Einstein of his theory of 
General Relativity), the rotating solution was found by Roy Kerr only 
in the sixties (Kerr 1963), which possibly tells us more on the lack of interest in the
field rather than on the mathematical difficulty of the problem (which was of course 
far from trivial).

All relevant General Relativity effects around a BH are scale invariant,
i.e. do not depend on the BH mass. It is therefore convenient to measure
all distances in units of the so-called gravitational radius, $r_g=GM/c^2$.
It is also useful to introduce the adimensional angular momentum per unit
mass, $a=Jc/GM^2$, called for simplicity ``spin'' here-in-after. 

In Boyer--Lindquist spherical coordinates (namely $t$, $r$, $\phi$, $\theta$, with
the usual meaning of symbols), the Kerr metric can be written as:

\begin{eqnarray}
\nonumber
ds^2 = -\left(1-{2r \over \Sigma}\right)dt^2 - \left({4ar\sin^2{\theta} \over \Sigma}\right)dtd\phi \\
\nonumber
+ \left({\Sigma \over \Delta}\right)dr^2   + \Sigma d\theta^2 \\
+ \left(r^2 + a^2 +{2a^2r\sin^2{\theta} \over \Sigma}\right)\sin^2{\theta}d\phi^2
\end{eqnarray}

\noindent
where:

\begin{equation}
\nonumber
 \Sigma = r^2 + a^2\cos^2{\theta}; ~~~~~~  
\Delta = r^2 - 2r + a^2
\end{equation}

\noindent
(If the Black Hole electric charge is not null, then, in geometrized units,
$\Delta = r^2 - 2r + a^2 + Q^2$). 
For $a$=0, the Schwarzschild metric is obtained:

\begin{eqnarray}
\nonumber
ds^2 = -\left(1-{2 \over r}\right)dt^2 + \left(1-{2 \over r}\right)^{-1}dr^2 \\
+ r^2\left(d\theta^2 + \sin^2{\theta}d\phi^2\right)
\end{eqnarray}

The radius of the Event Horizon, i.e. the surface of ``no return'', is given
by $R_{EO}=1 + \sqrt{1 - a^2}$. This implies that 0$\le$$a$$\le$1,  i.e. that
there is a maximum value for the spin. When $a$=1 the BH is said to be
maximally rotating; in this case the radius of the Event Horizon is 
equal to the gravitational radius, while it is 2$r_g$ (the ``Schwarzschild
radius'') for a static ($a$=0) BH.\footnote{It is important to recall here that Thorne (1974) 
has shown that in the standard accretion disc model the
radiation emitted by the disc and swallowed by the BH produces a counteracting
torque which limits the spin to a maximum value of $\sim$0.988, corresponding to 
$R_{EO}\sim$1.23.} It is interesting to note that the
Schwarz\-schild radius corresponds, in a pure Newtonian calculation,
to the radius a star should have in order than 
at its surface the escape velocity is equal to $c$.
Indeed, Black Holes (or invisible stars, as they were called at the time)
were predicted in this way more than
two centuries ago by Michell (1783) and Laplace (1796), even if of course
they could not imagine that from such objects nothing, not only the light, could escape. 

An important General Relativity  
effect is the gravitational redshift. Photons get out of the
gravitational potential of the BH only by losing energy, being therefore
redshifted. In Schwarzschild metric:

\begin{equation}
\nu_{obs}/\nu_{em}=\sqrt{1-{2 \over r}}
\end{equation}

\noindent
where $\nu_{em}$ and $\nu_{obs}$ are the emitted and observed (at infinity)
frequencies of the photon, and $r$ the emission radius 
(see also Fig.~\ref{all}). In Kerr metric, a similar
formula can be written only for the photons emitted on the rotation axis,
where it reads:  

\begin{equation}
\nu_{obs}/\nu_{em}=\sqrt{1-{2r \over r^2 + a^2}}
\end{equation}

\noindent
For
any other point, the ``dragging of the inertial frame'', i.e. the corotation
of the space-time with the BH spin makes gravitational and Doppler shifts
not separable.

\section{Line emission from accretion discs}

\subsection{General and historic remarks}

Accretion on Black Holes, at least for bright systems, it is widely 
believed to occur via an accretion disc, where gravitational energy
can be efficiently dissipated and eventually converted into
radiation. Accretion discs are very complicated systems, and the details
of the physical processes are far from be fully understood. 
For what follows, however, we only have to assume: that the disc is 
geometrically thin (i.e. its height is always much smaller than its radius
at any radius), so that it may be approximated with a thin slab on the
equatorial plane; that it is homogenous enough in order that clumpiness
does not affect much the line emissivity;
and that it is optically thick, so that iron line 
fluorescent emission can be efficient. (Even these assumptions, however, may 
be questionable and have indeed been questioned several times. This however
is not the place to discuss when and how the results are modified 
releasing one or more of them).

I also assume that the iron line is due to fluorescent emission
following illumination (and photo--ionization) of the accretion disc
by an external source of X-rays. George \& Fabian (1991) and Matt
et al. (1991) discussed in detail the properties of the fluorescent
line for neutral matter, while Matt et al. (1993a, 1996), Nayakshin \&
Kallman (2001) and A.C. Fabian, R.R. Ross and collaborators in a serie
of papers (Ross \& Fabian 2005, and references therein) 
discussed the case of ionized matter. 

GR effects on the radiation emitted by an accretion disc were first studied
by Cunningham (1975), while Fabian et at. (1989) and Chen et al. (1989) where the
first to model line emission from relativistic discs and compare calculations with
observations. Different groups (too many to be quoted here; further references
can be found in: Fabian et al. 2000; Reynolds \& Nowak 2003; Fabian \& Miniutti 2005;
Karas, this volume) have since then performed
calculations of line profiles under different assumptions and physical conditions, mainly 
stimulated by the {\it GINGA} discovery that iron lines are almost ubiquitous in the X-ray
spectra of AGN (e.g. Nandra \& Pounds 1994).
Models of line profiles from accretion discs are also present in the widely used {\sc XSPEC} 
software package (Arnaud 1996)\footnote{see also http://xspec.gsfc.nasa.gov/} for X-ray spectral fitting. For many years 
the only models available in {\sc XSPEC} 
were the {\sc diskline} (Fabian et al. 1989) and {\sc laor}
(Laor 1991) models. The {\sc diskline} model
is fast and reliable, but it is valid only in
Schwarzschild metric and it is somewhat inaccurate, in particular for small radii and 
large inclination angles, 
because of the straight line approximation for the photon geodesics. The {\sc laor} model
is valid only for a maximally rotating black hole; it is fast and reliable, but it is based
on a rather coarse grid of parameters. The limitations to these two codes were obviously
due to the limited power of the computers at the time they were written: more detailed fitting
codes would simply have been unmanageable. Computers improved a good deal since then, and
now fully relativistic
codes in Kerr metric (allowing for the entire range of spin)
have became publically available even if not yet included in the standard {\sc XSPEC}
release: the {\sc ky}  (Dovciak et al. 2004a,b) and {\sc kd} 
(Beckwith \& Done 2004, 2005) suits.\footnote{Just before submitting this contribution,
Brenneman \& Reynolds (2006) presented one new fully relativistic
code for spectral fitting in {\sc XSPEC}.}

\subsection{Relativistic discs and line emission}

The inner radius of the accretion disc cannot be  smaller than 
the Innermost Stable Circular Orbit (ISCO). This of course does not
mean that there is no matter at radii lower than the ISCO; simply, 
the matter must spiral in (see Krolik \& Hawley 2002 for different definitions
of the ``edge'' of the disc). 
The ISCO depends on the BH spin and on whether the disc is
co-- or counter--rotating with the BH (see Fig.~\ref{all}):

\begin{equation}
r_{\rm ISCO} = 3+Z_2 \pm \big[\left(3-Z_1)(3+Z_1+2Z_2\right)\big]^{1 \over 2}
\label{eqISCO}
\end{equation}

\noindent
where

\begin{eqnarray}
\nonumber
Z_1 = 1+(1-a^2)^{1 \over 3}[(1+a)^{1 \over 3}+(1-a)^{1 \over 3}] \\
\nonumber
Z_2 = (3a^2+Z_1^2)^{1 \over 2}
\end{eqnarray}

\noindent
The -- (+) sign applies to 
co-- (counter--) rotating discs. Indeed, the decrease
of the ISCO with $a$ (for a corotating disc) provides a method to measure 
the spin (see next section).

Motion of matter in accretion discs is supposed to be dominated by the 
gravitational potential of the BH, and then rotation to be Keplerian. 
Close to the BH the Keplerian velocity, $v_K$, becomes very large, reaching a 
significant fraction of the velocity of light. In the Locally Non-Rotating
Frame (LNRF), i.e. the reference frame ``rotating with the Black Hole''
(Bardeen et al. 1972), we have (see also Fig.~\ref{all}):

\begin{equation}
 v_K/c = \frac{r^2-2a\sqrt{r}+{a}^2}{\left(r^2+a^2-2r\right)^{{1 \over 2}}\left(r^{3/2}
 +{a}\right)}.
\label{eqkepl}
\end{equation}

\noindent
It is interesting to note that $v_K$ can be as high as almost half the
velocity of light, implying that the Doppler shift and boosting
 may be very prominent (Doppler boosting is the brightening/dimming of the
flux when the matter is approaching/receding. It is a Special Relativity
aberration effect due to the fact that  $I_\nu / \nu^3$ is a Lorentz invariant). 
At these velocities, Special Relativity corrections
of the Dop\-pler effect must be included, with the result that
transverse Dop\-pler effect (i.e. the redshift of photons when matter has only
a transverse component of the velocity) is by no means negligible.

In General Relativity, photon geodesics are no longer straight lines (the
so-called ``light bending''). In Sch\-warz\-schild metric they 
still lie on a plane, and therefore the equation of the orbit can be written
in terms of only two coordinates, the radius and the azimuthal angle, $\Phi$,
on the plane of the trajectory. The differential equation 
describing the orbit is (Misner et al. 1973):

\begin{equation}
{d^2u \over d\Phi^2} = 3u^2-u
\end{equation}

\noindent 
where $u=1/r$. 
In Kerr metric the orbits are fully tridimensional, and the equation of motion
much more complex (Carter 1968).
As a result of light 
bending, geodesics of photons emitted in the far side of the disc 
are strongly curved making the disc appears as ``bended'' towards 
the observer, with a sort
of sombrero-like shape (see e.g. Luminet 1979, 1992).

\begin{figure*}
\includegraphics[width=85mm,height=85mm]{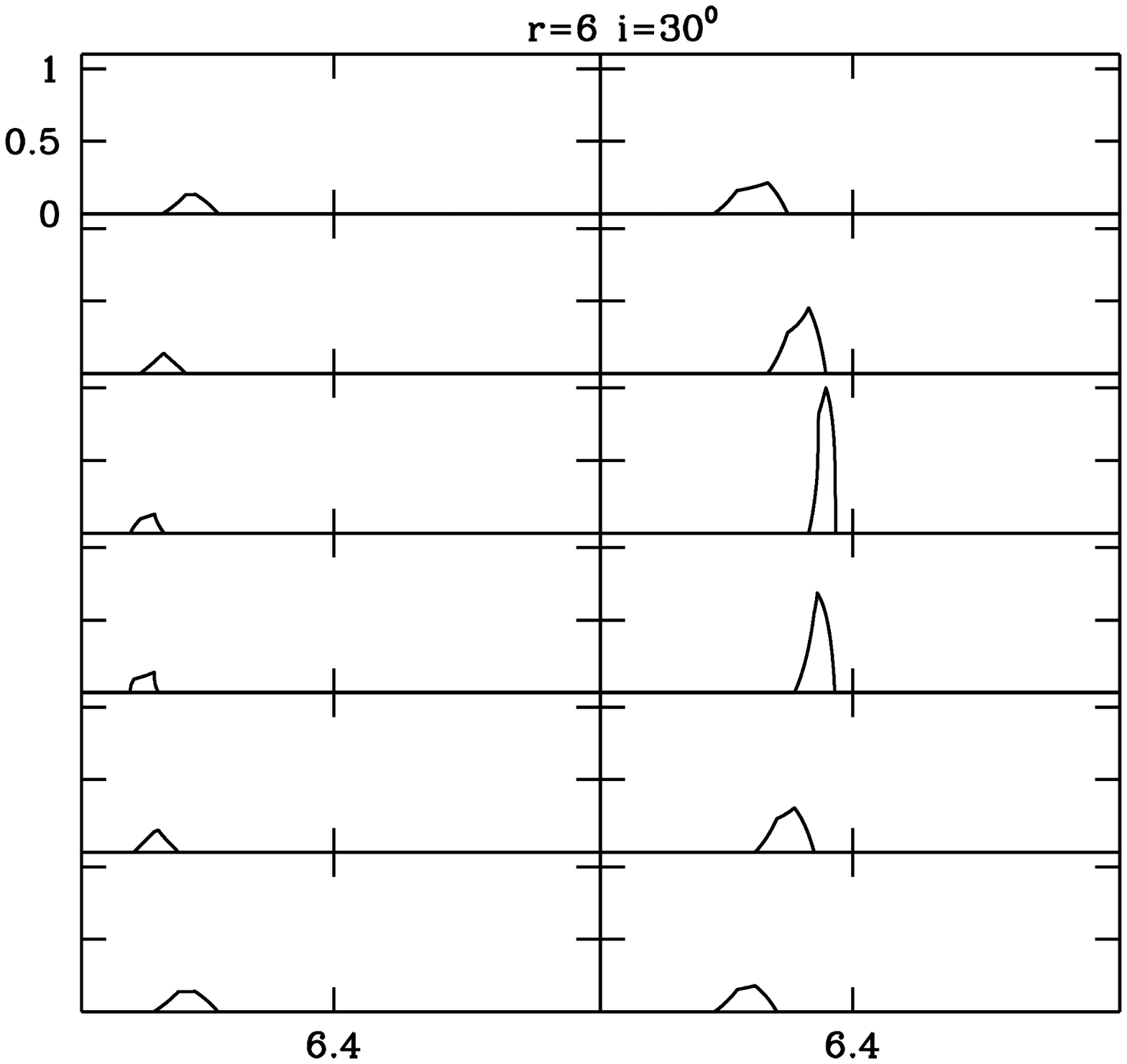}
\hfill
\includegraphics[width=85mm,height=85mm]{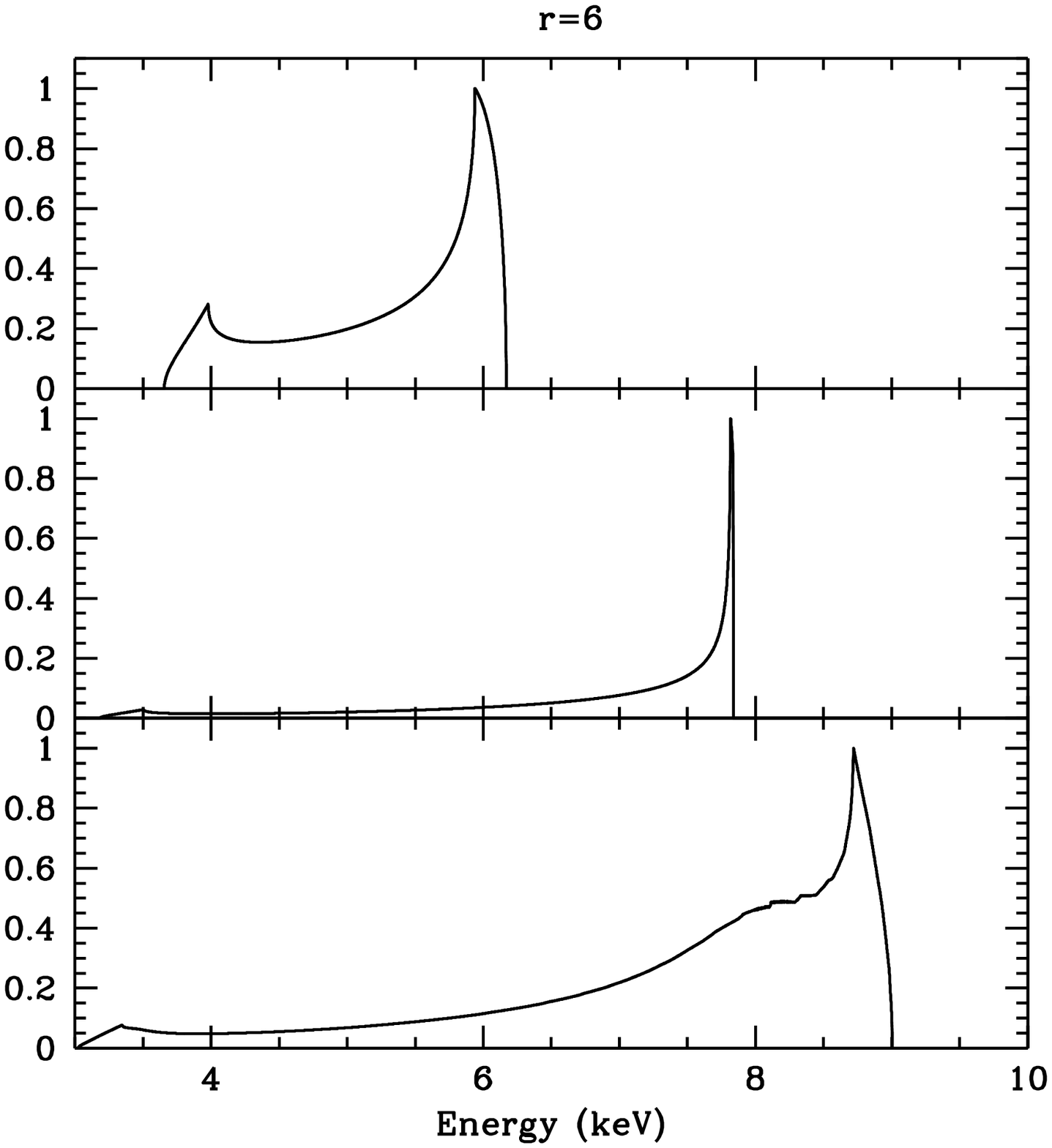}
\caption{{\it Left panel}: Iron line emission from an annulus at $r$=6, seen
with an inclination of 30$^{\circ}$. Each panel refers to an
azimuthal interval with $\Delta\phi$=30$^{\circ}$. The uppermost panel
on the left side refers to matter moving transversally on the near side
of the the disc ($\phi$=0$^{\circ}$). $\phi$ increases from top to bottom.
Left panels refer to receding matter, right panels to approaching matter. 
{\it Right Panel}: Azumuthally--averaged iron line profiles from a $r$=6 annulus, and for
30$^{\circ}$, 60$^{\circ}$ and 85$^{\circ}$ inclination angles (from top to
bottom.}
\label{prof_6_30}
\end{figure*}

\begin{figure*}[t]
\includegraphics[width=85mm,height=85mm]{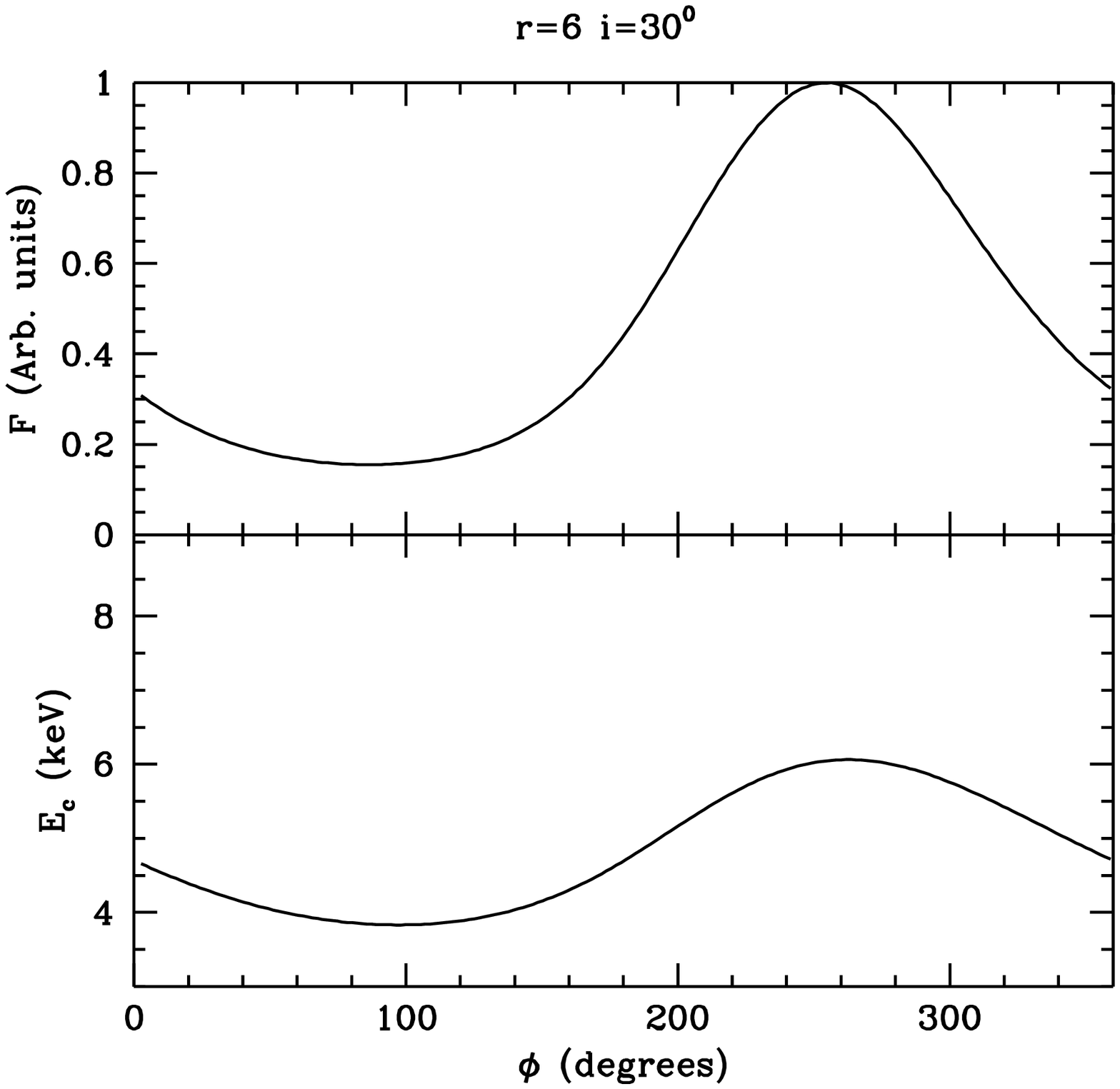}
\hfill
\includegraphics[width=85mm,height=85mm]{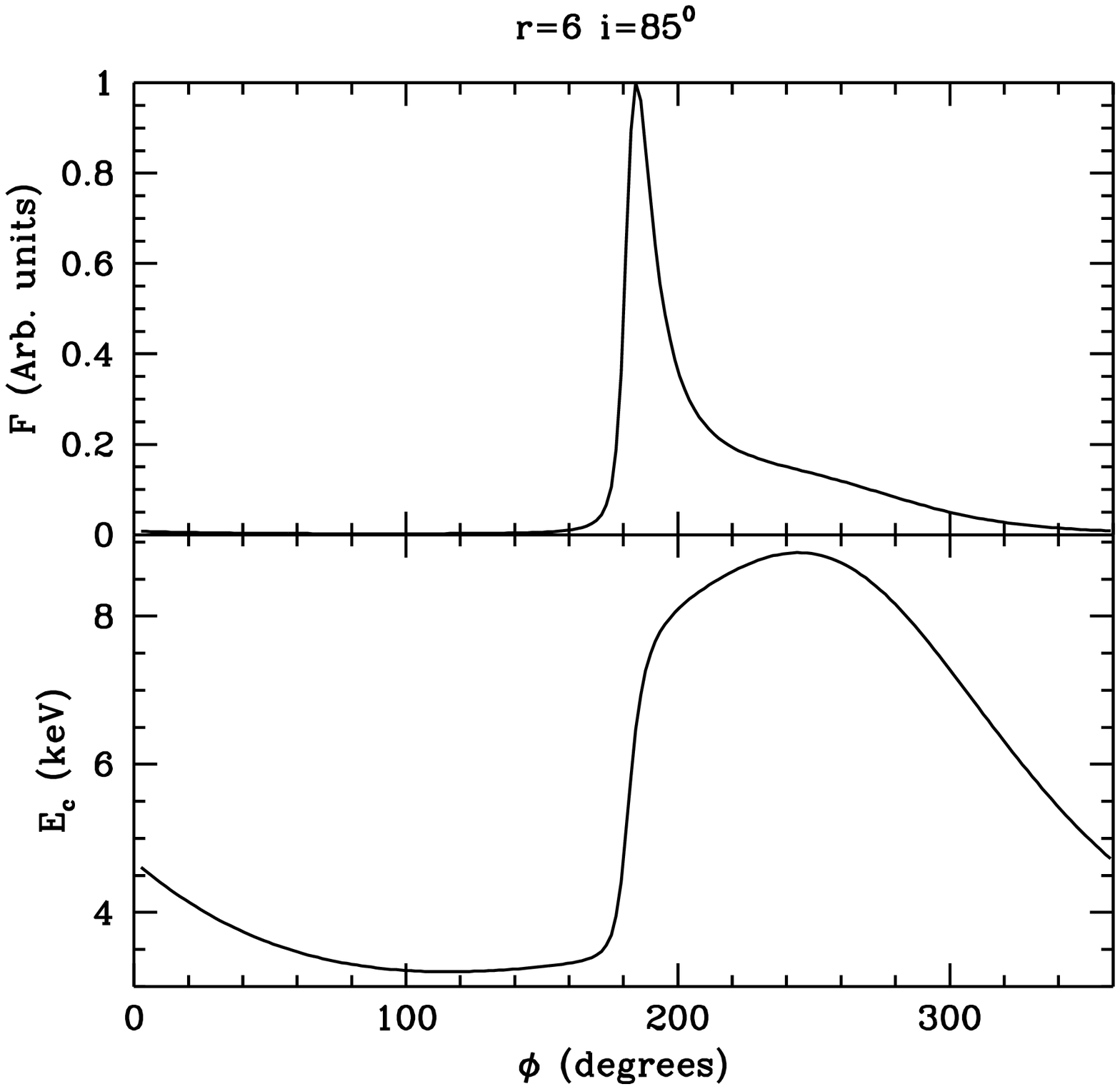}
\caption{{\sl Left panel}: Flux (upper panel) and centroid energy (lower panel) 
of the iron line as
a function of the azimuthal angle of emission. Radius ena inclination angle are
6 and 30$^{\circ}$, respectively. {\sl Right panel}: the same, but for an inclination
angle of 85$^{\circ}$.}
\label{orbit}
\end{figure*}

All these effects strongly modify the properties of emission lines. 
Let us for simplicity neglect natural
and thermal line broadening, so that the line profile, in the matter reference
frame, is a $\delta$-function. Given the topic of this conference, let us
also assume that the line is the neutral iron K$\alpha$ at 6.4 
keV.\footnote{This line is actually a doublet, with energies of 6.4055 and 
6.3916 keV and a branching ratio of $\sim$2:1 (Palmeri et al. 2003). 
As the broadening effects we are discussing here
are much larger than the $\sim$14 eV intrinsic separation, 
we will assume a single narrow line with a weighted mean energy of 6.4 keV).}
In Fig.~\ref{prof_6_30} (left panel) 
the line emission from an annulus at $r$=6, seen
with an inclination of 30$^{\circ}$, is shown after having divided the annulus
in azimuthal intervals, with $\Delta\phi$=30$^{\circ}$. The uppermost panel
on the left side refers to matter moving transversally on the near side
of the disc ($\phi$=0$^{\circ}$). 
Even in this case, when classic Doppler effect is null,
emission is significantly redshifted due to the combination of gravitational
and tranverse Doppler effects. Going down on the left side, matter starts 
receding and classic Doppler effect adds to further shift redwards the energy
of the photon, which is maximum for $\phi$=90$^{\circ}$, where instead the
flux is minimum due to Doppler (de--)boosting. In the right panels, matter
is instead approaching, but even for $\phi$=270$^{\circ}$, when the maximum
line-of-sight velocity is attained, emission is globally redshifted because
the Doppler effect (for so small radius and inclination angle) cannot fully
compensate for gravitational redshift. Due to Doppler boosting, flux is
maximum at $\phi$=270$^{\circ}$. The azimuthally averaged line profile is
shown in Fig.~\ref{prof_6_30} (right-upper panel) 
while the medium and lower right panels
show the  60$^{\circ}$ and  85$^{\circ}$ inclination angle cases, respectively.
Note that at higher inclinations the ``blue'' peak of the line profile is
actually blueshifted with respect to the rest frame energy, because of the
larger Doppler effect. Note also that, in the 85$^{\circ}$ case, structures
in between the red and blue peaks appear, due to the bending of the photons
emitted on the far side of the disc (e.g. Matt et al. 1993b). This effect is 
better illustrated in Fig.~\ref{orbit}, when the flux and centroid energy of the
line as a function of the azimuthal angle are shown for inclination angles of 30$^{\circ}$
and 85$^{\circ}$ ($r$=6). While in the 30$^{\circ}$ case the variations in flux are dominated
by the Doppler boosting, in the 85$^{\circ}$ case the peak of the emission occurs at  
$\phi\sim$180$^{\circ}$ due to the strong light bending. 

\begin{figure*}[t]
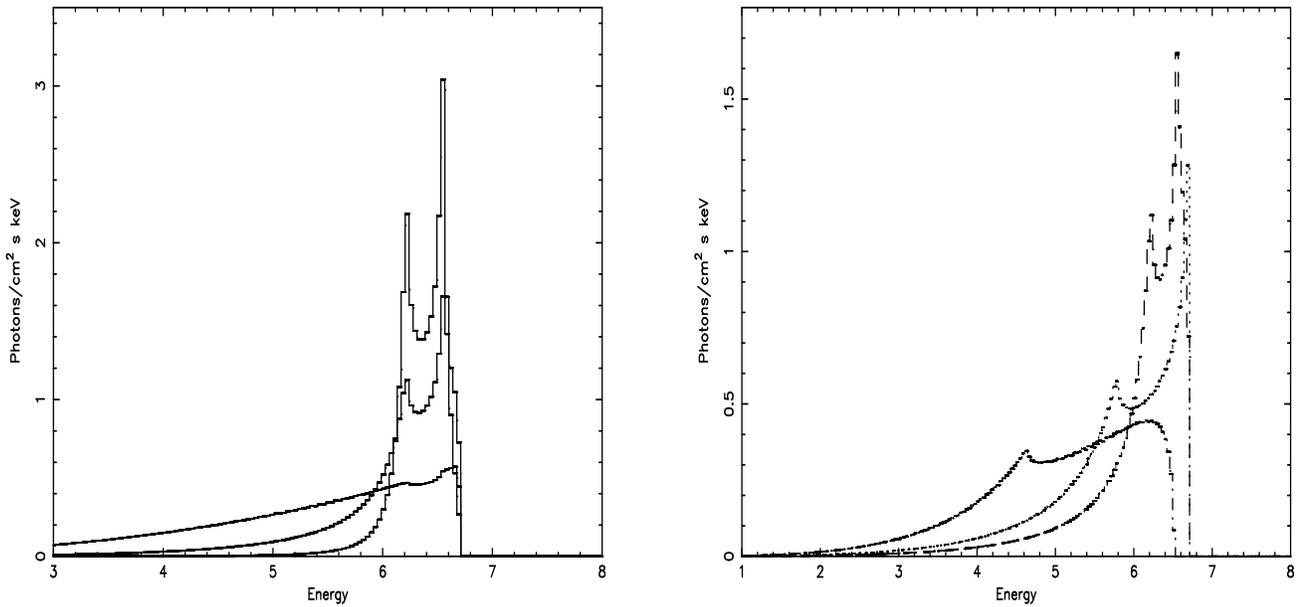

\includegraphics[width=80mm,height=80mm, angle=-90]{prof_emiss.ps}
\hfill
\includegraphics[width=80mm,height=80mm, angle=-90]{rout.ps}
\caption{{\sl Left panel}: iron line profiles for a maximally rotating BH, extending
from the ISCO to 400 $r_g$, with a 30$^{\circ}$ inclination angle.
Profiles refer to power law emissivity laws with $q$=1,2,3 (from top to
bottom). {\sl Right panel}: Iron line profiles for a maximally rotating BH, extending
from the ISCO to 400, 50 and 10 $r_g$ (from top to bottom) with a 30$^{\circ}$ inclination angle
and $q$=2. Here and in the next figure
profiles have been calculated with the code {\sc kyrline} (Dovciak
et al. 2004a,b) in the {\sc XSPEC} software package.} 
\label{prof_emiss}
\end{figure*}

Line profile is further modified by integration over the entire disc. To do that,
a crucial ingredient is the radial emissivity law, $\xi$. For the iron
fluorescent line, which is emitted following external illumination
(e.g. George \& Fabian 1991; Matt et al. 1991), $\xi$ depends mainly on the
geometry of the system. It is customary to assume a power law emissivity
law, $\xi \propto r^{-q}$. If $q<2$, the outer regions dominate the
emissivity, while the inner regions prevail for  $q>2$. Fig.~\ref{prof_emiss}
(left panel) show the impact on line profiles of different choices of $q$. 
The right panel
instead show, for a given value of $q$ (=2) the line profiles for different
values of the outer radius of the emitting region ($r_{\rm out}$=10, 50, 400 $r_g$).

Actually, the emissivity law is likely to be more complex than a simple 
power law. Even in the simplest case, the so-called ``lamp-post'' model
in which the primary emitting region is a small cloud on the BH axis
(as in aborted jet models, e.g. Ghisellini et al. 2004), the emissivity
law is, {\sl neglecting GR effects and radiative transfer subtleties}, 
given by: $\xi \propto (h^2+r^2)^{-{3 \over 2}}$, where $h$ is the height
of the emitting point; $\xi$ is then a power law ($q$=3) only for large
radii. Once the effects on the emissivity of the incident angle
(Matt et al. 1991) and, especially, of GR (light bending, gravitational
shift) are included, the emissivity is significantly modified (e.g. Martocchia \& Matt
1996, Martocchia  et al 2000, 2002). As noted by Martocchia et al. (2002), the
steep emissivity law found for the iron line emission of the Seyfert 1 galaxy
MCG--6-30-15, as observed by XMM--$Newton$ (Wilms et al. 2001), 
can be explained by this ``geometrical'' effect. A more general case,
in which the emitting region is no longer forced to stay on the BH axis,
has been studied by Miniutti et al. (2003, and this volume).

\section{Measuring the spin and mass of BHs}

Iron line profiles from relativistic accretion discs provides
potentially very powerful methods to measure the mass and the spin of
the Black Holes. Pros and cons of these methods are briefly discussed 
in the following paragraphs.

\subsection{Spin}

\begin{figure*}[t]
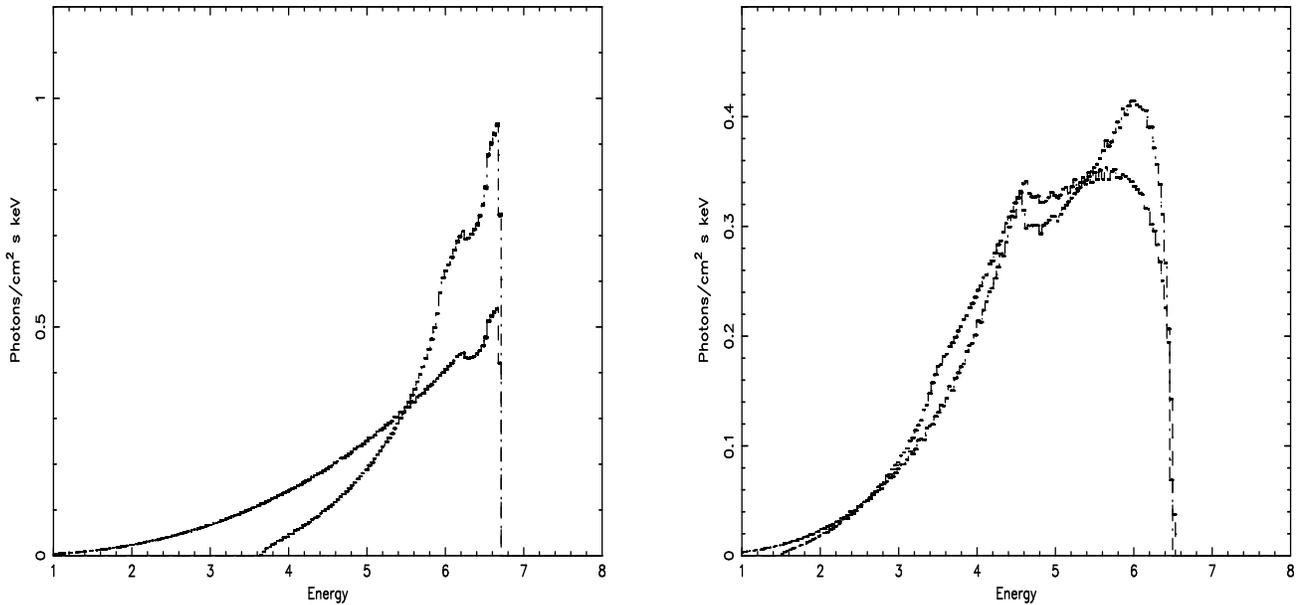

\includegraphics[width=80mm,height=80mm, angle=-90]{spin.ps}
\hfill
\includegraphics[width=80mm,height=80mm, angle=-90]{kerrnokerr.ps}
\caption{{\sl Left panel}: iron line profiles for a maximally rotating BH 
and a static BH, 
extending from the ISCO to 400 $r_g$, with a 30$^{\circ}$ inclination angle.
Note that the profile for the spinning BH extends to much lower energies. 
{\sl Right panel}: iron line profiles for a maximally rotating BH and a 
static BH, 
extending from $r$=2 to 10 $r_g$, with a 30$^{\circ}$ inclination angle, to
illustrate the differences due to the metric.} 
\label{spin}
\end{figure*}

Almost invariably, methods to measure the Black Hole spin make use, directly
or indirectly, of the dependence of the ISCO on $a$. Methods based on the iron
line make no exception. The smaller the inner disc radius, the lower (due to
gravitational redshift) the energy to which the profile extends.   
In Fig.~\ref{spin}, left panel, profiles from accretion discs
around a static and a maximally rotating Black Holes, 
in both cases extending down
to the ISCO, are shown. The advantage of this method is that it is 
very simple and straightforward, at least conceptually. Moreover, no
detailed physical modeling of the line emission is required: the spin
is measured from the low end of the profile, independently of the exact
form of the profile itself (which must be used only to break any degeneracy
with the inclination angle).

There are, however, also some limitations and caveats to this method that
must be considered. First of all, strictly speaking the method provides
only a lower limit to the spin, because the disc (or at least the iron line
emitting region) could in principle not extend down to the ISCO. Technically,
zero-intensity energies are far from trivial to be measured. Finally, 
even if the disc (properly said) stops at the ISCO, the region within
the ISCO (the so-called plunging region)
in not empty, and line emission may arise from the matter 
free-falling onto the Black Hole (Reynolds \& Begelman 1997; Krolik \& Hawley 2002), 
even if matter is expected to be significantly ionized there. 
 If the inner
radius results to be smaller than 2, there is of course no ambiguity. 
Otherwise, one could always rely to the subtle differencies in the
profiles due to the metrics themselves (Fig.~\ref{spin}, right panel) 
which, at least
for small radii, are after all not so negigible and will be hopefully exploited
by the next generation of large area X-ray satellites. 

\subsection{Mass (in Active Galactic Nuclei) }

Iron K$\alpha$ reverberation mapping of structures in the profile (Stella 
1990) or of integrated quantities (Equivalent Width, centroid energy and
width, Matt \& Perola 1992) has been suggested, in analogy with the method
routinely used for optical
broad lines, to measure the BH mass in AGN
(this technique is practically unapplicable in Galactic Black Hole
systems because of the very short time scales involved, and the 
much lower typical flux per light--crossing time). 
It is a conceptually simple but
technically very difficult technique. First of all, it requires a lot
of photons. Worst than that, the Transfer Function, which describes
how the line follows variations
of the illuminating continuum, is strongly geometry-dependent. With respect
to the BLR reverbaration mapping, one here has the advantage
that the geometry of the 
illuminated region can be assumed a priori (i.e. the accretion disc), but
has the disadvantage that the geometry of the illuminating region is unknown
(in the BLR case a point-like source is a safe assumption, given the much
larger distance of the illuminated matter). 

On the other hand, if the iron line is emitted in a small spot on the
accretion disc (corotating with the disc at the Keplerian velocity), the
BH mass could be easily and precisely measured, once the spot radius
is known (Dovciak et al. 2004c). Such a hot spot may
be due to a localized flare, possibly of magnetic origin, just above the
disc surface.
Because there is some evidence (albeit still controversial) for spot-like
emission in AGN 
(e.g. Turner et al. 2002; Dovciak et al. 2004c; Iwasawa et al. 2004;
Pechacek et al. 2005, and references therein), let us discuss this 
case in some detail. 

A spot on the accretion disc at a radius $r$ has an orbital period
(as measured by an observer at infinity) given by:

\begin{equation}
T_{\rm orb} = 310~\left(r^\frac{3}{2}+a\right)
M_7\quad\mbox{[sec]},
\label{torb}
\end{equation}

\noindent
where $M_7$ is the mass of the black hole in units of $10^7$ solar masses.
If the spot radius and the BH spin can be estimated, the measurement of the
orbital period immediately provides the Black Hole mass (note that the spin
is relevant only for small radii; when $r \gg 1$, when the spin is hard to 
measure, it fortunately becomes irrelevant).  

\begin{figure*}[t]
\includegraphics[width=80mm,height=80mm]{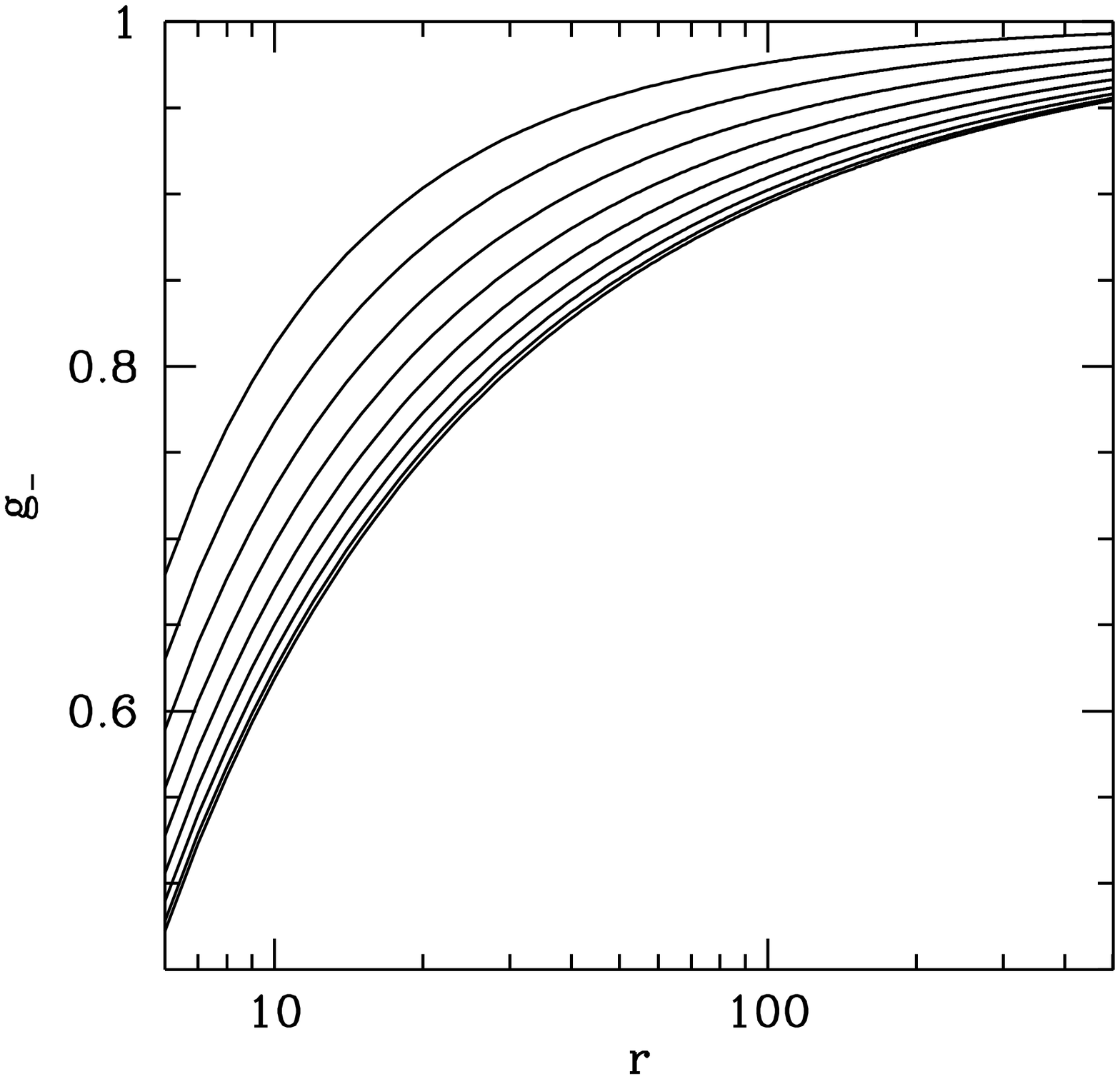}
\hfill
\includegraphics[width=80mm,height=80mm]{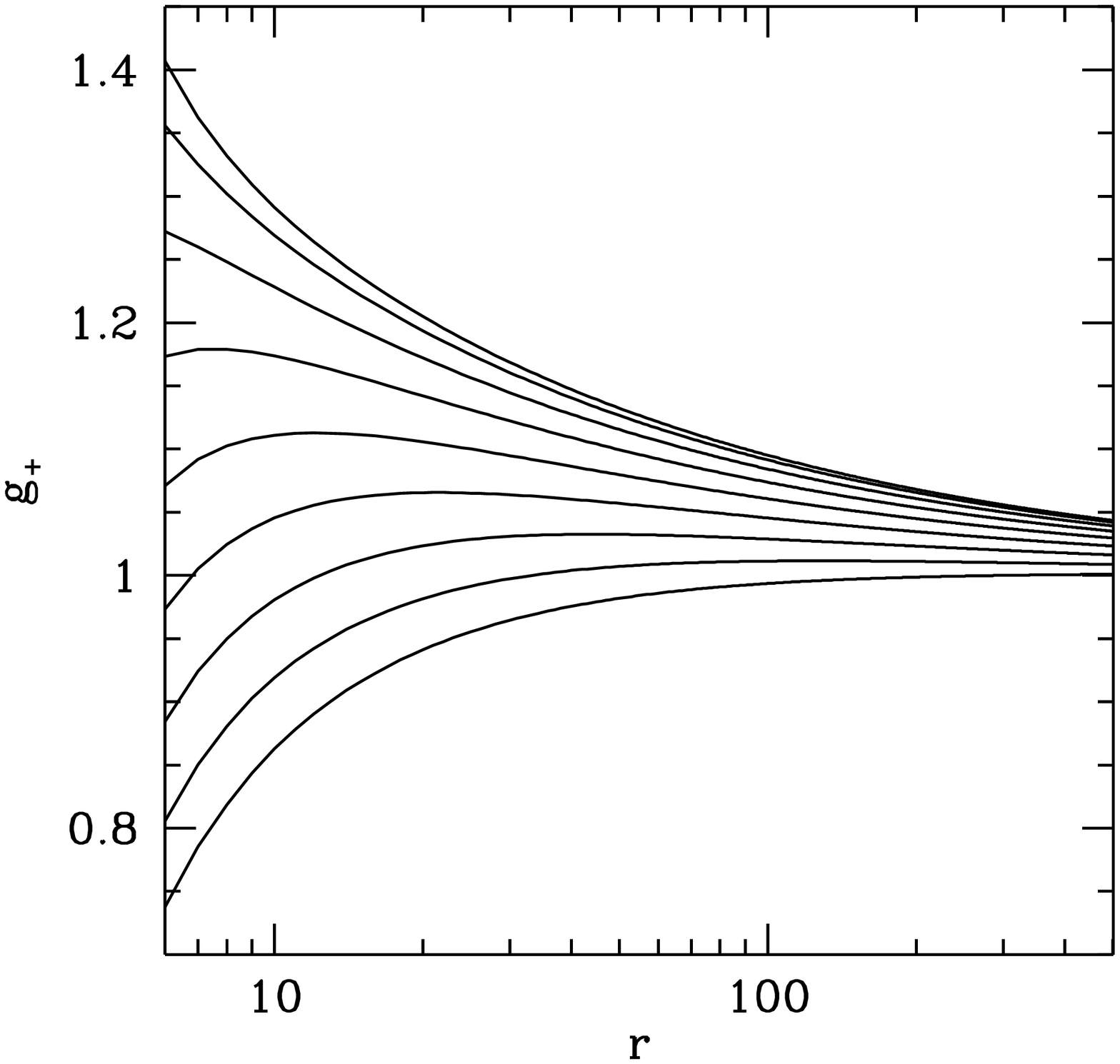}
\caption{The minimum ($g_{-}$, left panel) and maximum ($g_{+}$, right panel) shift factor
as a function of radius for 9 different values of the inclination angle: 
5$^{\circ}$, 15$^{\circ}$, .., 85$^{\circ}$ (from top to bottom in the left panel,
from bottom to top in the right panel). Calculations, which are for
Schwarzschild metric, are based on the
analytic, approximated formula (Eq.~\ref{gfac}) 
discussed in Pechacek et al. (2005). }
\label{gf}
\end{figure*}

In practice, it is well possible that, in low S/N spectra, only the blue peak
is visible (see Fig.~\ref{prof_6_30}), resulting in transient and relatively
narrow features. For low radii and inclination angles
the features may appear  
redshifted with respect to the rest frame energy. If only the blue peak is
visible, and then the entire profile cannot be reconstructed, it will be
impossible to tightly constraint the emission parameters, and only
allowed intervals for the radius and angle can be derived from the energy
shift. To this purpose, Pechacek et al. (2005) found a simple 
approximated formula which
gives, with a very good accuracy, the shift as a function of the radius and
the polar and azimuthal angle in the Schwarzschild metric. Calling $g$
the shift factor, i.e. the ratio between observed (at infinity) and
emitted energies, we have:

\begin{equation}
g = \frac{[r(r-3)]^{1/2}}{r+\left[r-2+
            4\left(1+\cos{\phi}\sin{\theta}\right)^{-1}\right]^{1/2}
            \sin{\phi}\sin{\theta}}
\label{gfac}
\end{equation}

\noindent
where $\phi$, as usual, is the azimuthal angle while $\theta$ is the polar
angle (i.e. the inclination angle in case of a disc).
 In Fig.~\ref{gf}, the maximum and minimum values of the shift factor
are shown as a function of the radius for different
inclination angles.

\section{Conclusions}

Iron lines are probably the best tools to probe GR effects in the vicinity
of Black Holes. Spectral distortions are much easier to study in lines
then in continua, because of their intrinsic narrowness - broadening can
be safely assume to arise mainly, when not exclusively, from such effects. 
Even if many important observational results have already been obtained
(as amply discussed in many papers in this volume), much is still to be
done, especially in using iron lines to estimate the mass and the spin of
the Black Hole. Indeed, relativistic iron lines are still a major scientific 
driver for next generation, large area X-ray satellites.

\acknowledgements
I wish to thank all my collaborators during the more than 15 years in which
I've been working in this field. For some of the plots in this paper I 
have made use of numerical codes developed
by M. Dovciak and V. Karas.

\newpage%%%%%%%%%%%%%%%%%%%%%%%%%%%%%%%%%%%%%%%%%%%%%%%%%%%%%%

\end{document}